\documentstyle[preprint,aps]{revtex}

\begin{document}

\draft

\tighten

\title{Quantum Kinetic Equations and Dark Matter Abundances Reconsidered}

\author{Anupam Singh\footnote{E--mail: \tt singh@lanl.gov}}

\address{Theoretical Division, T-8, Los Alamos National Laboratory,
Los Alamos, NM 87545}

\author{Mark Srednicki\footnote{E--mail: \tt mark@physics.ucsb.edu}}

\address{Department of Physics, University of California,
         Santa Barbara, CA 93106
         \\ \vskip0.5in}

\maketitle

\begin{abstract}
\normalsize{
Starting from a Caldeira-Leggett model for the interaction of a system with an
environment, Joichi, Matsumoto, and Yoshimura have reconsidered the derivation
of the quantum Boltzmann equation.  They find an extra term that accounts for
the effects of virtual particles, and which drastically changes the results for
relic densities of stable, weakly interacting massive particles (WIMPs), and
for the decay products of unstable particles.  We show, however, that this
modified Boltzmann equation does not properly account for the interaction
energy between the massive particles (which are decaying or annihilating) and
the thermal bath of light particles.  We argue that the conventional Boltzmann
equation gives the correct result.
}
\end{abstract}


Joichi, Matsumoto, and Yoshimura (hereafter JMY) \cite{jmy12,jmy3}
and Matsumoto and Yoshimura (hereafter MY) \cite{my}
have carefully reconsidered the derivation
of the quantum Boltzmann equation for heavy particles
embedded in a thermal bath of light particles.  JMY treat the case of 
unstable heavy particles, and MY treat
the case of a stable, weakly interacting massive particles
(WIMPs) that annihilate to light particles.  In both cases
they find new terms in the quantum Boltzmann equation that
account for the effects of virtual heavy particles, and that
drastically change the usual formula for the equilibrium abundance of these
particles.  In particular, the usual calculation of relic abundances of
WIMPs is completely changed, with the result that a WIMP with a mass
in excess of approximately $1\,$GeV is would overclose the universe,
for a broad range of interaction strengths with the light particles.

These surprising results must be taken seriously, since previous
derivations of the quantum Boltzmann equation (for weakly interacting
massive particles) can involve uncontrolled approximations and 
possibly arguable assumptions (see, e.g., \cite{kb} for a typical
treatment).  In this context the analyses of JMY and MY are among the most 
rigorous ones available.

The basic issue raised by JMY is more easily
understood in the context of an unstable, decaying particle (rather than
stable, annihilating particles).  
Given a spin-zero particle $\varphi$ with mass $M$ and energy 
$E({\bf p})=({\bf p}^2+M^2)^{1/2}$ at a temperature $T$,
the conventional formula for its equilibrium number density is
\begin{eqnarray}
n_\varphi &=& \int{d^3p\over(2\pi)^3}\,{1\over e^{E({\bf p})/T} - 1} 
\nonumber \\
       &=& \cases{ \zeta(3)T^3/\pi^2 & for $T \gg M$, \cr
                   (MT/2\pi)^{3/2}e^{-M/T} & for $T \ll M$. \cr }
\label{n}
\end{eqnarray}
JMY, on the other hand, argue that if the particle
is weakly coupled, and unstable with a decay width $\Gamma \ll M$,
then we should have instead \cite{jmy3} 
\begin{equation}
n_\varphi = \int{d^3p\over(2\pi)^3}
       \int_p^\infty d\omega\, 
       {\Gamma/2\pi \over \left(\omega-E({\bf p})\right)^2 + (\Gamma/2)^2}
       \,{1\over e^{\omega/T} - 1}\;.
\label{n2}
\end{equation}
That is, we should allow the energy of the unstable particle to
vary according to a Breit-Wigner function, rather than
be fixed at $\omega=E({\bf p})$.  (For simplicity of notation,
we have left out a time-dilation factor of $M/E({\bf p})$ that should
multiply $\Gamma$; this will not affect our subsequent analysis,
which is primarily concerned with the nonrelativistic regime.)
If we take the limit $\Gamma \to 0$, then the Breit-Wigner function
becomes $\delta(\omega-E({\bf p}))$, and we recover Eq.~(\ref{n}).  
On the other hand, if we take $T \ll \Gamma \ll M$, then 
the integral is dominated by the region near $\omega \sim T$, 
and we have instead \cite{jmy3}
\begin{eqnarray}
n_\varphi &=& {\Gamma\over 4\pi^3 M^2}\int_0^\infty d\omega\,
              {1\over e^{\omega/T}-1}\int_0^\omega dp\,p^2 
\nonumber \\
          &=&  {\pi\over 180}\,{\Gamma \over M^2}\, T^4
               \quad \hbox{for $T \ll \Gamma \ll M$.}
\label{n3}
\end{eqnarray}
This is drastically different than the usual result, Eq.~(\ref{n});
in particular, there is no exponential Boltzmann suppression.
We see that this is essentially because the $\varphi$ particles 
that are being counted in Eq.~(\ref{n3}) are far off shell, 
with energy near zero.

While Eq.~(\ref{n2}) may seem plausible, it leads to some surprising
conclusions.  Let us assume (following \cite{jmy12,jmy3}) that 
the $\varphi$ particle decays into two massless spin-zero $\chi$ particles
via an interaction
${\cal H}_{\rm int} = {1\over2}\mu\varphi\chi^2$; thus we have
$\Gamma = \mu^2/32\pi M$.  Now suppose that we place a hot gas of light 
$\chi$ particles in a large box, at a
temperature $T \ll M$.  
The number density of $\chi$ particles is
$n_\chi = \zeta(3)T^3/\pi^2$
and their energy density is $\rho_\chi = \pi^2 T^4/30$.
After thermal equilibrium is established
between $\varphi$ and $\chi$ particles, there should be
a number density $n_\varphi$ of $\varphi$ particles 
given by Eq.~(\ref{n3}).  The corresponding energy density 
$\rho_\varphi$ is obtained by including an extra factor of $E({\bf p})$
in the integrand of Eq.~(\ref{n2}) 
(and not, as one might guess, an extra factor of $\omega$).
For $T \ll M$, this implies
\begin{equation}
\rho_\varphi = M n_\varphi \sim \Gamma T^4/M \;.
\label{rhophi}
\end{equation}
We now see that the ratio of $\varphi$ energy density to $\chi$ energy density
is independent of temperature:
\begin{equation}
\rho_\varphi / \rho_\chi  \sim \Gamma / M \;.
\label{ratio}
\end{equation}
This strikes us as odd, since there would not seem to be a source
of virtual heavy $\varphi$ particles in the limit of zero temperature for
the massless $\chi$ particles. 

The situation worsens for the case of stable, annihilating $\varphi$
particles treated by MY \cite{my}.  For an interaction of the form
${\cal H}_{\rm int} = {1\over2}\lambda \varphi^2 \chi^2$, MY find
\begin{equation}
n_\varphi \sim \lambda(T/M)^{1/2}T^3 \;.
\label{n4}
\end{equation}
This implies $\rho_\varphi \sim \lambda (MT)^{1/2}T^3$, and hence
\begin{equation}
\rho_\varphi / \rho_\chi \sim \lambda (M/T)^{1/2} \;.
\label{ratio2}
\end{equation}
Thus, for $T \ll M$, we see that the energy density in
virtual heavy $\varphi$ particles greatly exceeds the energy density in
on-shell massless $\chi$ particles.  (This is not ruled out by energy
conservation; the original temperature of the $\chi$ gas would simply 
drop as the energy flows into virtual $\varphi$ particles.)
Eq.~(\ref{ratio2}) would seem to imply that (for example)
the cosmic microwave background radiation is accompanied by
a much larger energy density of virtual heavy particles.
We believe that this is not a tenable proposition.

Where, then, is the flaw in the MY analysis?  Consider 
a system coupled to an environment via an interaction,
\begin{equation}
H= H_{\rm sys} + H_{\rm env} + H_{\rm int} \;,
\label{h}
\end{equation}
where we assume that 
$H_{\rm sys}$ and $H_{\rm env}$ are positive semidefinite operators.
We wish to determine the energy of the system when it is 
in thermal equilibrium with the environment.
The most obvious candidate for this energy is
\begin{equation}
E_{\rm sys} = \langle H_{\rm sys}\rangle_T \;,
\label{e}
\end{equation}
where the angle brackets denote canonical thermal averaging
with subtraction of the zero-point energy,
\begin{equation}
\langle \ldots\rangle_T = {\mathop{\rm Tr}\ldots e^{-H/T}\over
                           \mathop{\rm Tr}       e^{-H/T}     }
                          -\langle 0|\ldots|0\rangle \;.
\label{t}
\end{equation}
This definition of $E_{\rm sys}$ is the one used by JMY and MY.
However, it is reasonable if and only if
\begin{equation}
\Bigl|\langle H_{\rm int}\rangle_T\Bigr| \ll \langle H_{\rm sys}\rangle_T \;.
\label{hh}
\end{equation}
If Eq.~(\ref{hh}) does not hold, then the interaction between
system and environment is effectively strong (no matter how small
the coupling may be), and the appropriate division between system
and environment is unclear.

The analyses of JMY and MY are based on consideration of a Caldeira-Leggett
model \cite{cl} of coupled harmonic oscillators, grouped into
terms according to Eq.~(\ref{h}).  We will show below that in this model,
at low temperature and weak coupling,
\begin{equation}
\langle H_{\rm int}\rangle_T  \simeq -2\langle H_{\rm sys}\rangle_T    \;.
\label{is}
\end{equation}
We see that the negative interaction
energy more than compensates for the system energy,
which our qualitative arguments indicated was much too large.

To demonstrate Eq.~(\ref{is}), we use the model presented in \cite{jmy3}.
A slightly different model was used in \cite{jmy12}; 
we have checked that Eq.~(\ref{is}) holds in the model of \cite{jmy12} as well.
The model of \cite{jmy3} is
\begin{eqnarray}
H_{\rm sys} &=& E_1\,c^\dagger c \;,
\label{sys} \\
H_{\rm env} &=& \int_{\omega_c}^\infty d\omega\,\omega\,
                b^\dagger(\omega)b(\omega) \;,
\label{env} \\
H_{\rm int} &=& \int_{\omega_c}^\infty d\omega\,\sqrt{\sigma(\omega)}\,
                \left[c^\dagger b(\omega) + b^\dagger(\omega)c \right] \;.
\label{int} 
\end{eqnarray}
Here $c$ and $b(\omega)$ are harmonic-oscillator operators 
with commutation relations
$[c,c^\dagger]=1$ and $[b(\omega),b^\dagger(\omega')]=\delta(\omega-\omega')$,
$\sigma(\omega)$ is a frequency-dependent coupling, 
and $\omega_c$ is a lower cutoff; we assume $\omega_c \ll E_1$.  
The exact solution of this model involves changing to new variables $B(\omega)$
such that 
\begin{equation}
H= H_{\rm sys} + H_{\rm env} + H_{\rm int} 
 = \int_{\omega_c}^\infty d\omega\,\omega\, B^\dagger(\omega)B(\omega) \;,
\label{h2}
\end{equation} 
where $[B(\omega),B^\dagger(\omega')]=\delta(\omega-\omega')$,
and the original operators are given in terms of the new ones via
\begin{eqnarray}
c &=& \int_{\omega_c}^\infty d\omega\,\sqrt{\sigma(\omega)}
                                      f(\omega)B(\omega) \; ,
\label{c} \\
b(\omega) &=& B(\omega) + O(\sigma) \;.
\label{b}
\end{eqnarray}
Here the function $f(\omega)$ is given by
\begin{equation}
f(\omega) = {1 \over \omega - E_1 + \Pi(\omega) + i\pi\sigma(\omega)} \;,
\label{f}
\end{equation}
where 
\begin{equation}
\Pi(\omega) = {\rm P}\int_{\omega_c}^\infty d\omega'\,
                     {\sigma(\omega')\over\omega'-\omega} \; .
\label{pi}
\end{equation}
The $O(\sigma)$ term in the formula for $b(\omega)$ will not be needed;
we will treat the coupling as weak, $\sigma(\omega) \ll E_1$,
and work to leading nontrivial order in $\sigma$.  
This means we can neglect $\Pi(\omega)$
compared to $E_1$, and treat $E_1$ as the renormalized single-particle
energy; this point is thoroughly discussed in \cite{jmy12,jmy3,my}.

We now wish to compute $\langle H_{\rm sys}\rangle_T$ and
$\langle H_{\rm int}\rangle_T$.  (We can also compute
$\langle H_{\rm env}\rangle_T$, but the result is infinite,
due to the infinite number of harmonic oscillators in the environment.)
This is entirely straightforward; the formula we need is
\begin{equation}
\left\langle B^\dagger(\omega')B(\omega)\right\rangle_T
 = {1\over e^{\omega/T}-1}\,\delta(\omega'-\omega) \;.
\label{tb}
\end{equation}
Using Eqs.~(\ref{sys},\ref{c},\ref{f},\ref{tb}), we have
\begin{eqnarray}
\langle H_{\rm sys}\rangle_T &=& E_1 \int_{\omega_c}^\infty 
                                     d\omega'\,d\omega\,
                                     \sqrt{\sigma(\omega')\sigma(\omega)}\,
                                     f^*(\omega')f(\omega)
                               \langle B^\dagger(\omega')B(\omega)\rangle_T
\nonumber \\
                             &=& E_1 \int_{\omega_c}^\infty 
                                     d\omega\,
                                     \sigma(\omega)|f(\omega)|^2
                                     \,{1\over e^{\omega/T}-1} 
\nonumber \\
                             &=& E_1 \int_{\omega_c}^\infty 
                                     d\omega\,
                                     {\sigma(\omega) \over
                                     (\omega-E_1)^2 + \pi^2\sigma^2(\omega)}
                                     \,{1\over e^{\omega/T}-1} \;.
\label{sys2}
\end{eqnarray}
We see the similarity with Eq.~(\ref{n2}).  At high temperature
and weak coupling,
the region near $\omega \sim E_1$ dominates, and we have
\begin{equation}
\langle H_{\rm sys}\rangle_T \simeq {E_1 \over e^{E_1/T}-1} \quad 
                               \hbox{for $\sigma(\omega) \ll T \sim E_1$.}
\label{sys3}
\end{equation}
This is the same result that one would obtain for a noninteracting oscillator.
On the other hand, at low temperature the low-$\omega$ region dominates, 
and we have
\begin{equation}
\langle H_{\rm sys}\rangle_T \simeq {1\over E_1}\int_{\omega_c}^\infty 
                                          d\omega\,
                                          {\sigma(\omega) \over
                                           e^{\omega/T}-1} \quad
                               \hbox{for $T \ll \sigma(\omega) \ll E_1$.}
\label{sys4}
\end{equation}

We now turn our attention to the interaction energy.  We begin by computing
\begin{eqnarray}
\langle c^\dagger b(\omega) \rangle_T &=& \int_{\omega_c}^\infty 
                                        d\omega'\,
                                     \sqrt{\sigma(\omega')}\,
                                     f^*(\omega')
                               \langle B^\dagger(\omega')B(\omega)\rangle_T
                                + O(\sigma)
\nonumber \\
                             &=& \sqrt{\sigma(\omega)}\,f^*(\omega)\,
                                 {1\over e^{\omega/T}-1}
                                + O(\sigma) \;.
\label{cdb}
\end{eqnarray}
{}From here on we do not display the $O(\sigma)$ correction.  We then have
\begin{eqnarray}
\langle H_{\rm int}\rangle_T &=&  \int_{\omega_c}^\infty 
                                     d\omega\,
                                     \sqrt{\sigma(\omega)}
        \left[\langle c^\dagger b(\omega)\rangle_T + \rm{c.c.}\right] 
\nonumber \\
                             &=& \int_{\omega_c}^\infty 
                                     d\omega\,
                                     { 2(\omega-E_1)\sigma(\omega) \over
                                     (\omega-E_1)^2 + \pi^2\sigma^2(\omega)}
                                     \,{1\over e^{\omega/T}-1} 
\label{int2}
\end{eqnarray}
At high temperature and weak coupling, we get 
\begin{equation}
\langle H_{\rm int}\rangle_T \simeq {\rm P} \int_{\omega_c}^\infty 
                                    d\omega\,
                                    {2\sigma(\omega)\over \omega-E_1}
                                    \,{1\over e^{\omega/T}-1} \quad
                               \hbox{for $\sigma(\omega) \ll T \sim E_1$.}
\label{int3}
\end{equation}
This is smaller than $\langle H_{\rm sys}\rangle_T$
due to a suppression factor of $\sigma(\omega)/E_1$.
Thus the interaction energy is small compared to the system energy,
as it should be.
If, however, we consider low temperature and weak coupling, then we get
\begin{equation}
\langle H_{\rm int}\rangle_T \simeq -\,{2\over E_1}\int_{\omega_c}^\infty 
                                          d\omega\,
                                          {\sigma(\omega) \over
                                           e^{\omega/T}-1} \quad
                               \hbox{for $T \ll \sigma(\omega) \ll E_1$.}
\label{int4}
\end{equation}
Comparing with Eq.~(\ref{sys4}) gives us Eq.~(\ref{is}).

Clearly, then, the proper identification of the system energy becomes a key
issue.  We do not have a definitive resolution of this puzzle.  However,
we offer the following comments regarding the specific problem of
WIMP relic densities.

It seems to us that the WIMP relics that survive to the present day
(and constitute the nonbaryonic dark matter) must be on-shell particles.
If they are to be virtual, they must be produced via processes involving
the present microwave-background photons; as discussed above, we find
this to be untenable.  Therefore, the correct computation is, we believe,
of the number density of {\it on-shell} particles.  
This leads us back to the usual Boltzmann equation, 
which assumes all particles are on-shell.
In fact, among the clearest derivations of this equation is the one
presented by MY, before they go on to consider virtual effects.
The usual Boltzmann equation of course leads to the standard result for
the relic WIMP density, which we believe is correct.

\begin{acknowledgments}

We thank John Ellis, Toby Falk, and Keith Olive for collaboration
on the early stages of this work.  We also thank Salman Habib and Emil Mottola
for helpful discussions.  This work was supported in part by the United States
Department of Energy at Los Alamos National Laboratory, by the National Science
Foundation through grant PHY--97--22022, and by the Institute of Geophysics and
Planetary Physics through grant 920.

\end{acknowledgments}

\end{document}